# Gate-tunable multiband transport in ZrTe$_5$ thin devices

Yonghe Liu[1,2#], Hanqi Pi[1,2#], Kenji Watanabe[3], Takashi Taniguchi[4], Genda Gu[5], Qiang Li[5,6], Hongming Weng[1,2], Quansheng Wu[1,2*], Yongqing Li[1,2*], Yang Xu[1,2*]

[1]Beijing National Laboratory for Condensed Matter Physics, Institute of Physics, Chinese Academy of Sciences, Beijing 100190, China

[2]School of Physical Sciences, University of Chinese Academy of Sciences, Beijing 100049, China

[3]Research Center for Functional Materials, National Institute for Materials Science, 1-1 Namiki, Tsukuba 305-0044, Japan

[4]International Center for Materials Nanoarchitectonics, National Institute for Materials Science, 1-1 Namiki, Tsukuba 305-0044, Japan

[5]Condensed Matter Physics and Materials Science Department, Brookhaven National Laboratory, Upton, New York 11973-5000, USA

[6]Department of Physics and Astronomy, Stony Brook University, Stony Brook, New York 11794-3800, USA

*Correspondence to: quansheng.wu@iphy.ac.cn; yqli@iphy.ac.cn; yang.xu@iphy.ac.cn

[#]These authors contributed equally: Yonghe Liu, Hanqi Pi.

**Abstract: Interest in ZrTe$_5$ has been reinvigorated in recent years owing to its potential for hosting versatile topological electronic states and intriguing experimental discoveries. However, the mechanism of many of its unusual transport behaviors remains controversial, for example, the characteristic peak in the temperature-dependent resistivity and the anomalous Hall effect. Here, through employing a clean dry-transfer fabrication method under inert environment, we successfully obtain high-quality ZrTe$_5$ thin devices that exhibit clear dual-gate tunability and ambipolar field effects. Such devices allow us to systematically study the resistance peak as well as the Hall effect at various doping densities and temperatures, revealing the contribution from electron-hole asymmetry and multiple-carrier transport. By comparing with theoretical calculations, we suggest a simplified semiclassical two-band model to explain the experimental observations. Our work helps to resolve the long-standing puzzles on ZrTe$_5$ and could potentially pave the way for realizing novel topological states in the two-dimensional limit.**

The Zirconium pentatelluride ZrTe$_5$ is a layered van der Waals material and has been shown to host distinct topological states in the vicinity to a topological phase transition (TPT). In the monolayer form, it is proposed to be a large-gap quantum spin hall insulator (QSHI)[1], whereas compelling evidence is still lacking. There are many novel physical properties that have been discovered for the ZrTe$_5$ crystals, ranging from high electron mobilities (up to ~ 640,000 cm$^2$V$^{-1}$s$^{-1}$)[2], the chiral anomaly[3], three-dimensional quantum Hall effect[4], and various parameter driven TPT[5-11]. Among the characteristic transport properties of ZrTe$_5$, the peculiar resistance peak at a temperature $T_p$ ~140 K was first observed about 40 years ago, concomitant with a sign reversal of the Hall and Seebeck coefficients happening across $T_p$ [12]. However, later studies show contrasting behaviors with the peak occurring in a wide temperature range from ~ 0 and 180 K[4,13,14]. The exact value of $T_p$ is highly dependent on the growth condition of the bulk crystals and can vary sample to sample. The origin of such resistance anomalies has been debated for a long time but still remained unclear. The formation of a charge density wave[15,16], a polaronic model[17,18], a semimetal-semiconductor transition[19], a Lifshitz transition[20,21], or a TPT[6,7,10] has been proposed.

Meanwhile, we have noticed that $T_p$ is generally smaller for ZrTe$_5$ single crystals grown by the flux method comparing with that grown by the chemical vapor transport (CVT) method[4,14]. For some samples grown by the flux method (nearly stoichiometric), semiconducting behavior down to the lowest temperature has been observed ($T_p \approx 0$)[14,22-26]. Another typical observation is a p-type Hall/Seebeck coefficient for $T > T_p$ and a n-type/multi-carrier behavior for lower temperatures ($T < \sim T_p$). It has been suggested that the amount of Te vacancies, which act as electron donors, could contribute to the different transport behaviors in the narrow-gap semiconductor ZrTe$_5$ [26]. However, without the ability to finely control the Te deficiency or charge doping, solid conclusions cannot be reached.

In-situ electrostatic gating has been widely used in van der Waals (vdW) materials to tune their charge densities and facilitate the understanding of a plethora of intriguing phenomena. In the early efforts of fabricating the thin gating devices of ZrTe$_5$, a main obstacle is the degradation of crystal quality during the sample preparation. High densities of p-type carriers and low mobilities (~$10^{19}$ cm$^{-3}$ and <~$10^3$ cm$^2$V$^{-1}$s$^{-1}$) are typically observed[27,28]. Meanwhile, efficient gating can hardly be achieved[23,28-30]. Here, we adopt a clean layer-by-layer dry-transfer fabrication method developed for vdW heterostructures in recent years (see Methods for more details). Thin ZrTe$_5$ flakes are mechanically exfoliated from the bulk crystals grown by the flux method[3]. The exposure to air or solvents are fully avoided during the device fabrication process, which helps maintain the intrinsic properties of the ZrTe$_5$ crystals. We also encapsulate the ZrTe$_5$ on both sides with hexagonal boron nitrides (hBN), which further improve the sample quality and prevent sample degradation during loading into the cryostats[31].

In this study, electrical transport measurements with changing gate voltages, temperatures, and magnetic fields are performed. Unless otherwise specified, the data presented below are collected from one representative device (about 12 nm thick, with an optical image shown in Fig. 1b). It exhibits clear ambipolar field effects and realizes efficient dual-gate tunability with mobilities up to ~15,000 cm$^2$V$^{-1}$s$^{-1}$. The resistance peaks, occurring at temperature $T_p$, can be systematically controlled by the gate voltages. The $T_p$ reaches its minimum near the charge neutrality and increases on electron and hole-doped sides asymmetrically. The Hall measurements suggest that the resistance peak does not need to correlate with the sign change of Hall resistance on the hole-doped side. The anomalous Hall-like responses and further analyses reveal a trivial multiband origin and at least two bands contributing to the transport. The conclusion is well supported by the first-principles calculations. We hence understand that the finite $T_p$ is a natural result due to temperature dependent scattering time and chemical potentials in degenerately doped narrow-gap semiconductors.

We first introduce the dual-gated electric field effect measured at $T = 1.7$ K in Fig. 1. The longitudinal sheet resistance $R_{sq}$ ($R_{sq} = \frac{L}{W} R_{xx}$, with $L$ and $W$ being the channel length and width, respectively) at zero magnetic field (Fig. 1c upper panel) and anti-symmetrized Hall resistance $R_{xy} = \frac{R_{xy}(B = 1\ \text{T}) - R_{xy}(B = -1\ \text{T})}{2}$ at a magnetic field $B = 1$ T (Fig. 1c lower panel) are presented as functions of the top and bottom gate voltages ($V_{tg}$ and $V_{bg}$) in color contour plots. The $R_{sq}$ reaches a global maximum at small values of ($V_{tg}$, $V_{bg}$) = (0.40 V, 0.40 V), which corresponds to the chemical potential tuned to the charge neutrality point (CNP). It is also evidenced by the sign reversal of the Hall resistance near the same point.

The linecuts along the off-diagonal dashed white lines (corresponding to equal densities induced by the two gates) of the two color maps in Fig. 1c are shown in the upper panel of Fig. 1d. The device exhibits graphene-like ambipolar electrical field effect with CNP at $V_D = 0.40$ V and a large on/off ratio about 40[32, 33]. Both the conductance $\sigma_{xx} = 1/R_{sq}$ and the inverse of Hall resistance $1/R_{xy}$ increases linearly on the p-doped ($V_{tg} < V_D$) and n-doped ($V_{tg} > V_D$) sides of $V_D$, indicating effective tunning of the charge transport and carrier densities (Fig. 1d lower panel). The Hall mobility can be extracted by $\mu_H = \sigma_{xx} R_H^l$, where $R_H^l \approx R_{xy}/B$ is the low-field Hall coefficient for $B < 0.5$ T. We obtain $\mu_H = $ ~8,000 cm$^2$V$^{-1}$s$^{-1}$ and ~5,000 cm$^2$V$^{-1}$s$^{-1}$ for the n-type and p-type charge carriers, respectively. The quantum oscillations due to the Landau level formation have also been observed and shown in Fig. S7 of SI, indicating high quality of our device. While in early attempts of obtaining thin ZrTe$_5$ flakes, the device mobility is usually much lower and the gate tuning is inefficient (ionic gating[29] or SiO$_2$ gate[23, 28, 30]). Our results indicate that the hBN encapsulation greatly helps maintain the high quality of ZrTe$_5$ crystals and prevent sample degradation that can hardly be avoided by other fabrication methods in previous experiments[28-30]. Note that the gate voltages are applied symmetrically to maintain $C_{tg}(V_{tg}-V_D)=C_{bg}(V_{bg}-V_D)$ throughout the following part of the paper while only the $V_{tg}$ is labeled for simplification.

With the gate-tunability established, we now attempt to resolve the mystery of "resistance peak" by performing temperature dependent measurements at controlled densities. The curves of $R_{sq}$(T) at different gate voltages (selected along the dashed line in Fig. 1c) are plotted in Fig. 2, displaying resistance maxima at temperatures denoted as $T_p$ (highlighted by the vertical short markers). In general, the device exhibits insulating behaviors ($\frac{dR_{sq}}{dT} < 0$) at $T > T_p$ and metallic behaviors ($\frac{dR_{sq}}{dT} > 0$) at $T < T_p$. When the chemical potential is tuned close to the CNP ($V_{t(b)g} = V_D = 0.40$ V), $R_{sq}(T)$ keeps increasing upon cooling and $T_p$ is nearly suppressed to the lowest temperature. A thermal activation gap $\Delta \sim 40$ meV is extracted from the Arrhenius fit of $R_{sq} \sim \exp(\frac{\Delta}{k_B T})$ at the CNP (see Fig. S6 in SI). When the gate voltages are tuned away from the CNP, $T_P$ shifts to higher temperatures and the peak resistance deceases, as guided by the grey dashed curve in the electron-doped side ($V_{t(b)g} > V_D$) and red dashed curve in the hole-doped side ($V_{t(b)g} < V_D$). The trend reminds us of the temperature-dependent resistivity observed in ZrTe$_5$ bulk crystals grown by different methods. The $T_P$ typically falls into a range between 120 -180 K for the bulk samples grown by the chemical vapor transport (CVT) method[4-7, 9, 12, 13, 20, 26, 30, 34-44], while being smaller than 100 K for the flux-method grown samples (some have $T_p \approx 0$)[2-4, 14, 21-26, 43, 45-48]. The flux samples are believed to have less Te vacancies and smaller densities comparing to the CVT samples. Surprisingly, the feature of $R_{sq}(T)$ in different bulk samples can be realized in one single gate-tunable device here. We also observe asymmetry in the gate dependence of the resistance peaks on the two sides away from the CNP. On the hole doped side with increasing the doping density, the $T_p$ shifts to higher temperature in a faster fashion and the resistance peak decreases more substantially.

In order to better understand the above features, we have performed Hall measurements at different doping levels and temperatures. The temperature induced evolution of the Hall resistances at three representative gate voltages, $V_{tg}$ = -0.75 V (hole-doped side), 0.40 V (charge neutrality), and 1.20 V (electron-doped side), are shown in Fig. 3a. Negative slopes of the Hall resistance are always observed at $V_{tg}$ = -0.75 V, regardless of the temperature being below or above $T_p$ (~200 K, see the green curve in Fig. 2), indicating the hole dominance of the charge transport. In contrast, at $V_{tg}$ = 0.40 V and 1.20 V, stronger nonlinearity has been observed for the Hall resistances. With increasing temperature, the low-field ($B \sim 0$ T) Hall coefficient ($\frac{dR_{xy}}{dB}$) changes sign, followed by another "sign reversal" of the high-field ($B \sim 9$ T) Hall coefficient at higher temperatures.

We extract the two Hall coefficients (denoted as $R_H^l$ and $R_H^h$, respectively) and plot them as functions of temperature for the three gate voltages in Fig. 3b. Examples of extracting $R_H^l$ (slope of the red dashed lines) and $R_H^h$ (slope of the grey dashed lines) at $T$ = 150 K are shown in the insets. Again, smaller inconsistency between the $R_H^l$ and $R_H^h$ is observed in the hole-doped side ($V_{tg}$ = -0.75 V) and they only decrease (in magnitude) slightly with increasing temperature. While at $V_{tg}$ = 0.40 V and 1.20 V, the low-field ($B \sim 0$ T) Hall coefficient $R_H^l$ quickly changes sign at ~50 K and ~120 K, respectively. The high-field Hall coefficient $R_H^h$ is typically a measure of the total carrier density of the system (in the semiclassical transport limit of $\mu B \gg 1$, with $\mu$ being the effective carrier mobility), supported by its nearly constant value at low temperatures (< ~150 K). We note that the "sign reversal" of $R_H^h$, which is extracted near 9 T, does not necessarily indicate a change in the total carrier density at high temperatures (150 ~ 200 K, where $\mu$ is less than ~200 cm$^2$/Vs). The reasoning is further explored in SI. We notice that at the electron-doped side ($V_{tg}$ = 1.20 V), the resistance peak happening at $T_p \approx$ ~120 K (see Fig. 2) seems to correlate with the sign reversal of the low-field Hall coefficient ($R_H^l$), consistent with that observed in bulk crystals[4, 12, 19-21, 26, 37]. However, on the hole-doped side, the resistance peaks still emerge even without any sign change of the Hall resistances.

The nonlinearity of the Hall resistance observed in ZrTe$_5$ has been previously attributed to the anomalous Hall effect (AHE)[22, 23, 25, 49-51], the multi-band transport[2, 7, 14, 21, 26, 27, 29, 36, 37, 52], or the thermally excited two carriers [4, 18, 53]. Step-like Hall resistances can be obtained by subtracting $R_H^h B$ from the total Hall resistances at $T$ = 1.7 K (Fig. 4a), resembling the anomalous Hall effect observed in soft ferromagnetic conductors[54]. Although, the presence of AHE can give rise to the nonlinearity of the Hall resistance, we note that it is unlikely to explain the observation in our devices. The majority of recent experimental and theoretical studies suggest ZrTe$_5$ to have a bulk band gap (varying from ~6 -100 meV) [20, 35, 43, 55-58]. In one scenario, the AHE can result from the formation of Weyl nodes, which are sources and sinks of Berry flux, with the inversion of the conduction and valence bands (CB and VB) upon applying external magnetic fields[22, 51]. It requires the Zeeman energy $g\mu_B B$ ($g$ is the Landé g-factor) to exceed the band gap. Taking a relatively large $g \approx 21$ [37] and the estimated energy gap from $2\Delta \approx 80$ meV (twice the activation gap), we can approximately evaluate the critical magnetic field to be larger than 65 T, which is far beyond the value we applied here.

Another mechanism of generating non-zero Berry curvature is from the magnetic-field-induced spin splitting of the massive Dirac bands (schematics shown in the inset of Fig. 4b)[23, 25, 49, 50]. It gives rise to the AHE which saturates when the chemical potential only crosses one spin-split band. Hence the saturation field $B_0$ is Fermi energy ($E_F$) dependent. Based on the calculated energy dispersion of bulk ZrTe$_5$, we estimate $B_0$ from $B_0 \propto E_F \propto n^{2/3}$ (illustrated by the grey curve in Fig. 4b)[50], which exhibits strong density dependence and quickly becomes larger than ~30 T for the carrier density $n > \sim 10^{18}$ cm$^{-3}$. Experimentally, the saturation field $B_0$ can be extracted from the crossing point of the two dashed lines (tracing the low-field and high-field Hall slopes, respectively) at different doping levels (see examples in Fig. 4a), showing a trend of saturation with much smaller values as the chemical potential moves away from the CNP (Fig. 4b). The observation in our experiment is apparently inconsistent with the AHE induced from Zeeman splitting of the massive Dirac band.

The other likely cause of nonlinearity in the Hall signal is the multiband transport[37, 52]. As shown in Fig. 4c, we attempt to use the semiclassical two-band model to fit the experimental Hall curves ($T = 1.7$ K) at different total densities (details in SI)[14, 21, 26, 27, 36, 52]. This rather simplified model gives reasonably good fitting results (as shown by the dashed curves) for almost all the Hall resistances. The extracted carrier densities ($n_1$, $n_2$, and $n_1 + n_2$), mobilities ($\mu_1$ and $\mu_2$), and corresponding ratios ($n_2/n_1$ and $\mu_1/\mu_2$) are plotted as functions of the total density $n$ in Fig. 4d-4f. The transport can then be classified into the contribution from one band with lower densities and higher mobilities ($n_1$ and $\mu_1$), and the other band with higher densities and lower mobilities ($n_2$ and $\mu_2$). The electron mobility of band 1 is as high as ~15,000 cm$^2$V$^{-1}$s$^{-1}$. The $n_1$ and $n_2$ (black and red symbols in the top panel of Fig. 4d) are almost linearly tuned by the electrostatic gates, both changing their sign near the CNP. Their sum $n_1 + n_2$ (blue symbols) shows general agreement with the total carrier density $n$ calculated from the geometric capacitances and the gate voltages, supporting the validity of our fittings. The difference of $\mu_{1,2}$ at $n > 0$ and $n < 0$ indicates electron-hole asymmetries. The $n_2/n_1 > 1$ indicates band 2 is heavier than band 1 (effective mass $m^*_{1,2} \propto n_{1,2}^{2/3}$) and the difference is more apparent in the electron-doped side (see Fig. 4f).

The multiband nature of ZrTe$_5$ is also revealed by our first-principles calculations (Fig. 5a and 5b, see more details in Methods) as well as some other studies[25, 26, 52]. We obtain a ~70 meV bandgap, near which multiple electron and hole pockets can be identified (Fig. 5b). The shapes of the bands near the band edges are also very different for the CB and VB, contributing to the different transport properties (electron-hole asymmetry). Comparing with the above results from the two-band fittings of the Hall curves, the band 1 with higher mobilities and smaller effective masses should arise from the Dirac-like band near the $\Gamma$ point, while the band 2 represents the averaged contribution from the other bands. We have also calculated the resistivity as a function of temperature at a few fixed total densities using the BoltzTrap code[59]. For simplicity, we only consider the scattering time $\tau$ dominated by the electron-phonon interaction in which $\tau$ is proportional to $1/T$ and assume five times larger $\tau$ for the holes (the assumption is validated as shown by our scattering analyses in SI). As shown in Fig. 4c, the main resistivity behaviors can be reproduced, such as the resistivity peak and its shifting with increasing the electron or hole densities[26]. The temperatures for the resistivity peaks are plotted in Fig. 5d, showing general consistency with the experimental $T_p$ values extracted from Fig. 2. The small deviation should result from the oversimplified model, in which we only considered electron-phonon as the main scattering source and assume constant ratio of the scattering times between electron and hole bands. The dispersion of $T_p$ divide the $T$ - $n$ map into three regions, corresponding to the p-type metal ($T < T_P$ and $V_{tg} < V_D$), n-type metal ($T < T_P$ and $V_{tg} > V_D$), and p-type semiconductor ($T > T_P$), respectively. The schematic electronic dispersion with two sets of bands and the corresponding Fermi energies are illustrated for each region. Upon increasing the temperature, the chemical potential $\tilde{\mu}(T)$ approaches the intrinsic Fermi level, which is inferred to be closer the VB maxima due to the lighter effective masses.

We also note that the occurrence of the resistivity peak is a natural result in degenerate semiconductors with narrow gaps. The essence can be captured by a simplified model with even only considering one set of conduction and valence bands (detailed calculation and discussion can be found in SI). Further considering the difference in scattering time in the CB and VB will reproduce the $T_p$ difference in for the electron and hole-doped sides (Fig. S11). The temperature induced change of $\tilde{\mu}(T)$ is also in good agreement with the results reported previously: it tends to shift upward [35, 60] from the VB or shifts downward [19, 20, 61] from the CB when the temperature increases. In conventional metals with large Fermi energy, the chemical potential $\tilde{\mu}$ does not change appreciably since $k_B T \ll E_F$. However, in degenerate semiconductor with narrow-gap or semimetals, where the VB maximum and CB minimum are in proximity (few tens of meV) to the chemical potential, significant shifting of the chemical potential with temperature can happen [53, 62, 63].

In summary, we have investigated the multiband properties of ZrTe$_5$ thin flakes by systematic gating and temperature dependent transport measurements. The temperature of the resistance peaks ($T_P$) and nonlinearity of Hall resistance can be efficiently tuned by the gate voltages, through which we identify a multiband origin and the highly asymmetric CB and VB dispersions with narrow bandgaps. Though exotic mechanisms like TPT or interaction effects cannot be fully excluded for the resistance anomaly in bulk ZrTe$_5$ [6, 7, 11, 18], our results provide a crucial step toward understanding its peculiar transport behaviors and paves the way for further exploring topological properties in thin ZrTe$_5$ devices.

**Note**

The authors declare no competing financial interest.

**Acknowledgement**

This work was supported by the National Natural Science Foundation of China (Grants No. 12174439, 11961141011, 12274436, 12188101, and 11925408), the Strategic Priority Research Program of Chinese Academy of Sciences (Grant No. XDB28000000 and XDB33000000), the National Key Research and Development Program of China (Grant No. 2021YFA1401300, 2022YFA1403403, 2018YFA0305700, and 2022YFA1403800), the Informatization Plan of the Chinese Academy of Sciences (Grant No. CAS WX2021SF-0102), and the Innovation Program for Quantum Science and Technology (Grant No. 2021ZD0302400). Work at Brookhaven National Laboratory was supported by the U.S. Department of Energy, Office of Science, Office of Basic Energy Sciences, under Contract No. DE-SC0012704. The growth of hBN crystals was supported by the Elemental Strategy Initiative of MEXT, Japan, and CREST (JPMJCR15F3), JST.

**Supporting Information**

Further details of the device fabrication, transport measurement setup, theoretical calculation methods, magnetotransport data analysis, scatterings mechanisms, two-band model for fitting the Hall resistances, temperature-dependent resistivity from a simplified narrow-gap model, and additional datasets for supporting the conclusion in the main text.

**Figure**

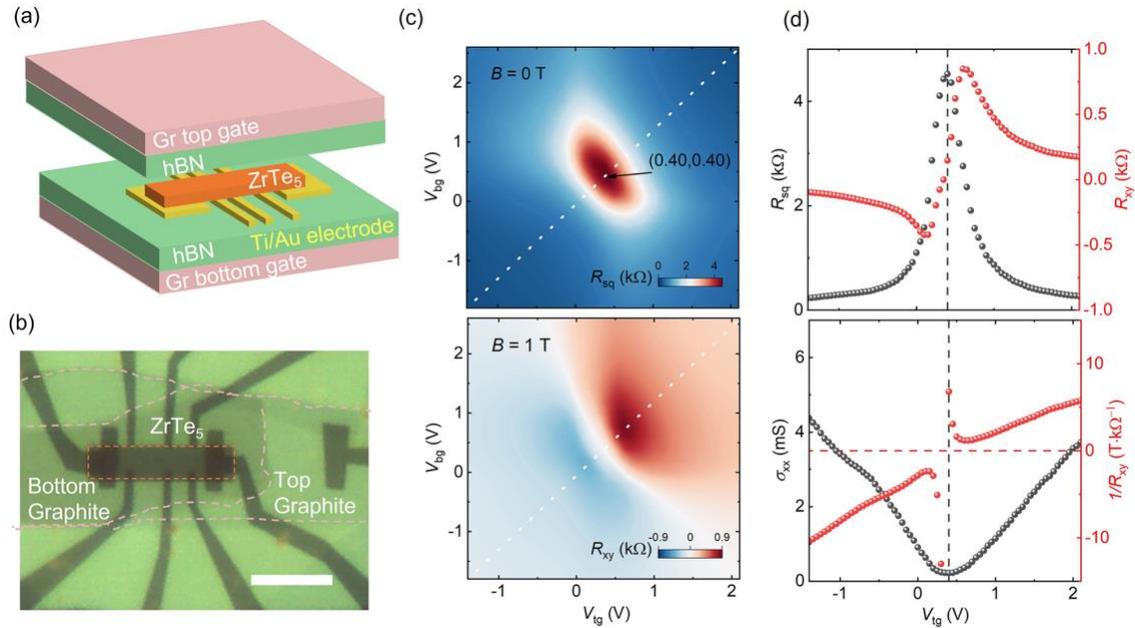

**Fig. 1. Device structure and dual-gated field effect of a thin ZrTe₅ device.** (a, b) Schematic and optical micrograph of the dual-gated thin ZrTe₅ (with thickness about 12 nm) Hall-bar device. The scale bar in (b) is 10 μm. (c) Color contour plots of the longitudinal sheet resistance $R_{sq}$ (upper panel $B = 0$ T), and Hall resistance $R_{xy}$ (lower panel, antisymmetrized at $B = \pm 1$ T) as functions of $V_{tg}$ and $V_{bg}$ measured at $T = 1.7$ K, with the corresponding linecuts along the dashed lines (equal density induced by each gate) presented in the upper panel of (d). The arrow in the upper panel of (c) and the vertical dashed lines in (d) highlight the charge neutrality point (CNP). The lower panel in (d) shows the conductance $\sigma_{xx} = 1/R_{sq}$ at $B = 0$ T and the inverse of the Hall resistance $1/R_{xy}$ at $B = 1$ T. Note that the x-axis is labelled as $V_{tg}$ while the $V_{bg}$ (not shown here) is simultaneously tuned according to the dashed line in (c) (same notation is used for the following figures).

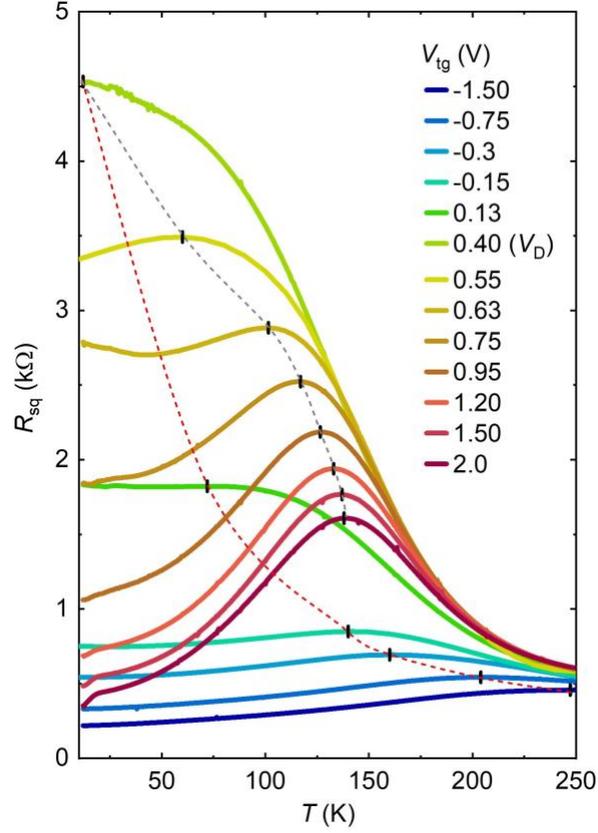

**Fig.2. Temperature-dependent resistances and scatterings at a few densities.** The $R_{sq}(T)$ shows pronounced electron-hole asymmetry. The short vertical lines mark the temperature ($T_P$) of the resistance peaks at different gate voltages. The gray and red dashed curves serve as guides to the eye for the tendency of $T_P$ at n-doped ($V_{tg} > 0.40$ V) and p-doped ($V_{tg} < 0.40$ V) sides, respectively. The data at $V_{tg}$ =-0.30, -0.15, and 0.55 V are extracted from dual-gated resistance mappings at different temperatures.

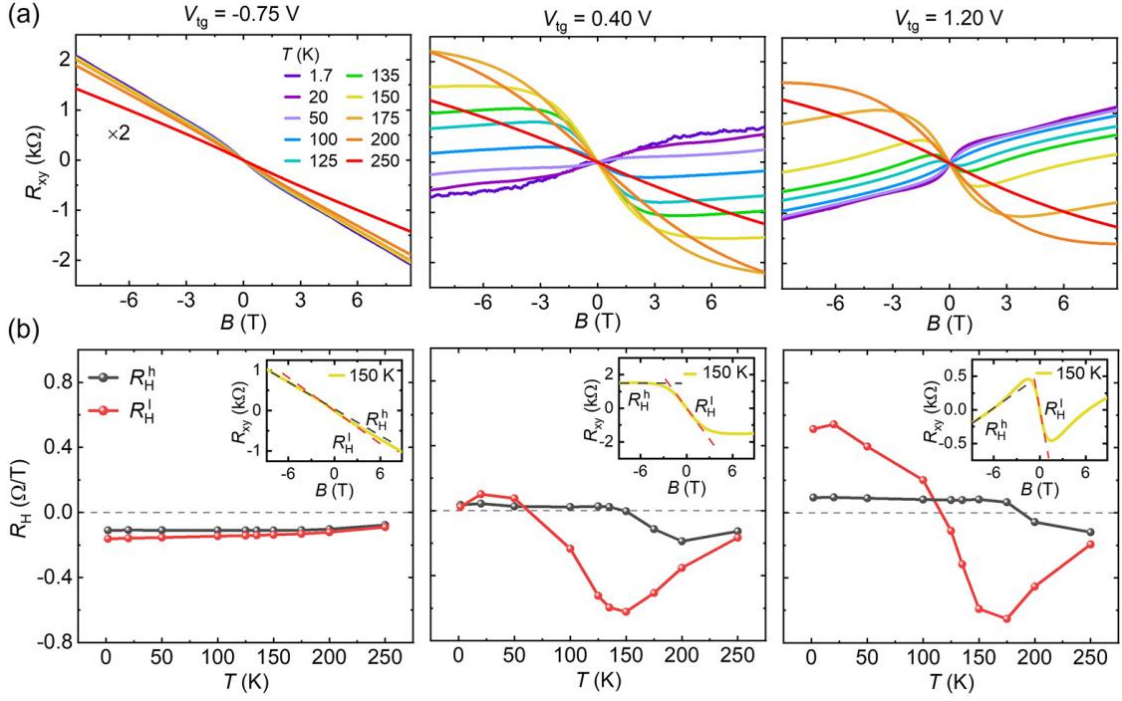

Fig. 3. **Temperature dependence of the Hall measurements at three representative gate voltages.** (a) Temperature-dependent evolution (from 1.7 to 250 K) of the Hall resistance $R_{xy}$ measured at $V_{tg}$ = -0.75 V (hole-doped side), 0.40 V (charge neutrality), and 1.20 V (electron-doped side), respectively. (b) The corresponding Hall coefficients $R_H^l$ ($R_H^h$) extracted from (a) at low (high) magnetic fields and plotted as functions of temperature. Examples of extracting $R_H^l$ (slope of red dashed line) and $R_H^h$ (slope of gray dashed line) at $T$ = 150 K are shown in the insets. All the plots share the same vertical-axis labeling on the left. The $R_{xy}$ is multiplied by a factor of two for clarity in the left panel of (a).

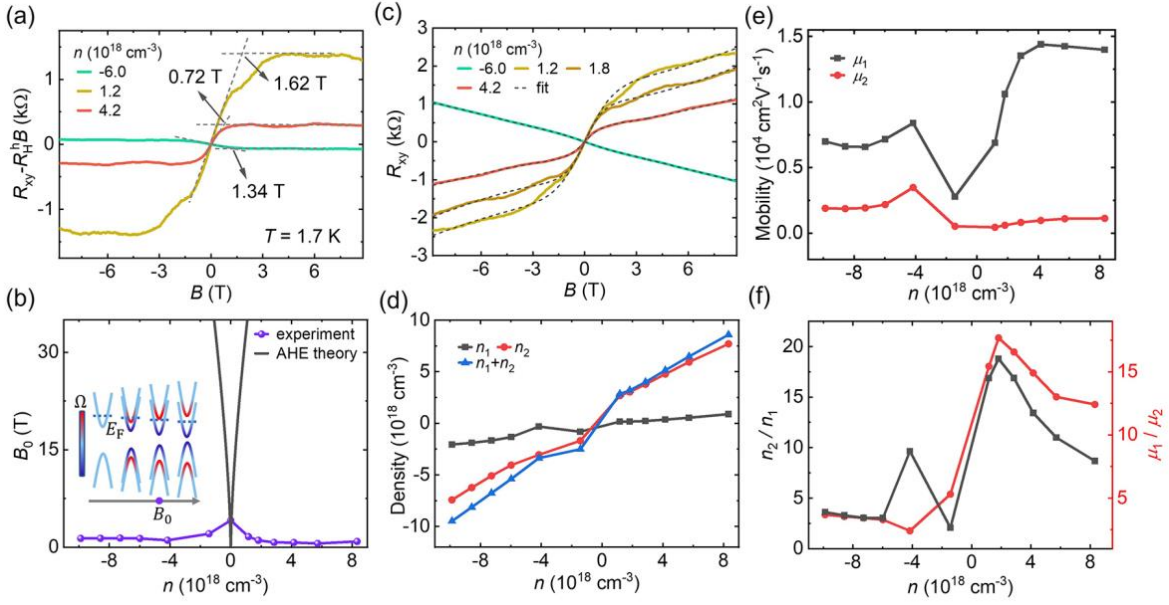

Fig .4. **Multiband induced anomalous-Hall-like response at 1.7 K.** (a) The extracted $(R_{xy} - R_H^h B)$ versus $B$ fields for three representative densities (corresponding to $n$ = -6.0, 1.2 and 4.7 ×10$^{18}$ cm$^{-3}$, respectively), showing anomalous-Hall-like features. We can define a "false" saturation field $B_0$ at the crossing point of the two dashed lines (tracing the low-field and high-field trends, respectively). (b) The carrier density dependence of $B_0$ (connected purple symbols), extracted using the method described above. The solid gray curve is the theoretically expected $B_0$ for the AHE from the Zeeman-split massive Dirac bands (schematic shown in the inset), assuming a relatively large Landé $g$-factor = 21. (c) The Hall resistance $R_{xy}$ (solid curves) and corresponding two-band-model fittings (dashed curves) at different gate voltages. (d, e) The carrier densities $n_i$ (d) and mobilities $\mu_i$ (e) extracted from the two-band-model fittings of the Hall resistances (i = 1, 2). (f) The corresponding ratios ($n_2/n_1$ and $\mu_1/\mu_2$) plotted as functions of the total density $n$.

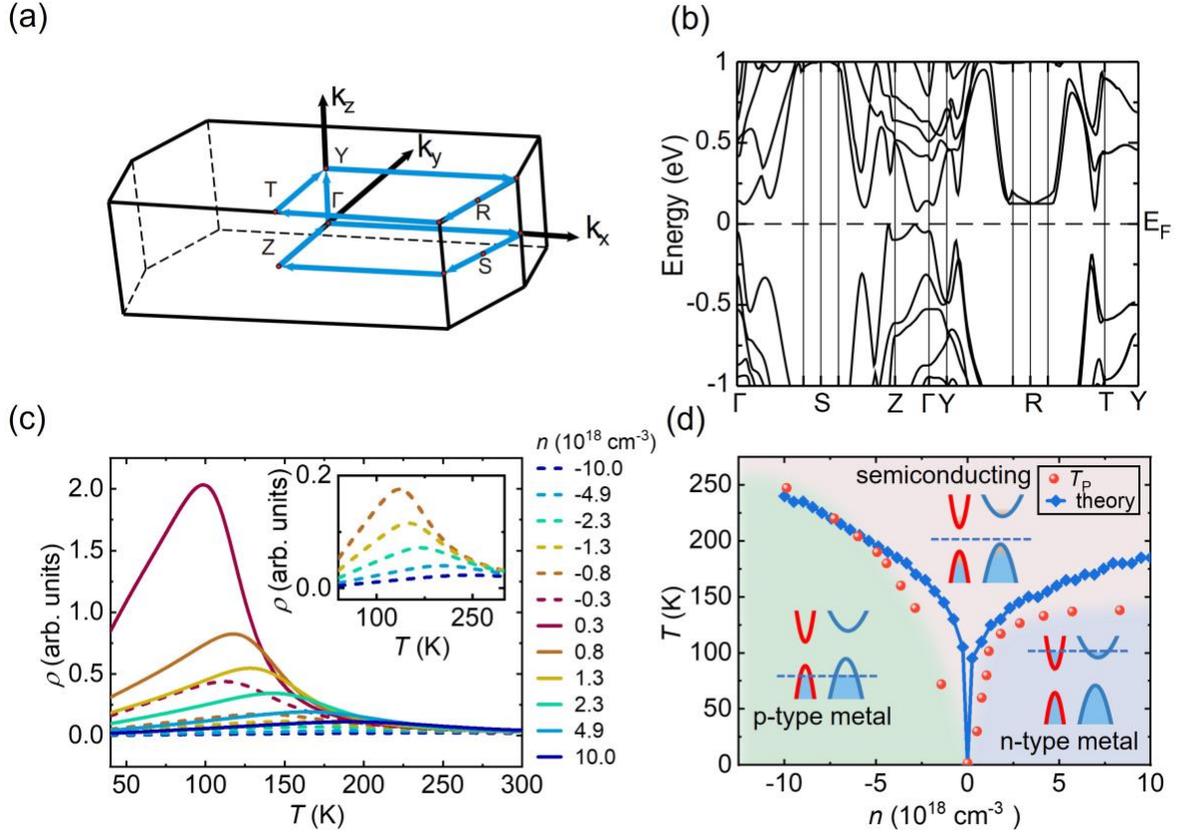

Fig. 5. **Theoretically calculated band structure and transport properties.** (a) The first Brillouin zone with the path (blue arrows) illustrated for the band structure calculation. (b) The band structure for bulk ZrTe$_5$. Note the multiband feature and very different structures of the valence bands and conduction bands. (c) The calculated resistivity $\rho$ as a function of temperature using the Boltztrap code at several carrier densities $n$. The scattering time $\tau$ is assumed five times larger for holes comparing to electrons. The inset shows a zoom-in plot of several most p-doped curves. (d) "Phase diagram" for the ZrTe$_5$ device. The red sphere symbols denote the experimentally extracted $T_P$, which divides the diagram into three regions: lower-temperature p-type and n-type metallic regime (showing $\frac{dR_{sq}}{dT} > 0$) and higher-temperature semiconducting regime (showing $\frac{dR_{sq}}{dT} < 0$). The boundary is also outlined by the theoretically obtained $T_p$ values (connected bule symbols) extracted from (c). Each region is illustrated with the schematic band structures and corresponding chemical potentials. The red and blue curves denote the lighter and heavier bands, respectively.

# Supporting Information

# Gate-tunable multiband transport in ZrTe$_5$ thin devices

Yonghe Liu[1,2#], Hanqi Pi[1,2#], Kenji Watanabe[3], Takashi Taniguchi[4], Genda Gu[5], Qiang Li[5,6], Hongming Weng[1,2], Quansheng Wu[1,2*], Yongqing Li[1,2*], Yang Xu[1,2*]

[1]Beijing National Laboratory for Condensed Matter Physics, Institute of Physics, Chinese Academy of Sciences, Beijing 100190, China

[2]School of Physical Sciences, University of Chinese Academy of Sciences, Beijing 100049, China

[3]Research Center for Functional Materials, National Institute for Materials Science, 1-1 Namiki, Tsukuba 305-0044, Japan

[4]International Center for Materials Nanoarchitectonics, National Institute for Materials Science, 1-1 Namiki, Tsukuba 305-0044, Japan

[5]Condensed Matter Physics and Materials Science Department, Brookhaven National Laboratory, Upton, New York 11973-5000, USA

[6]Department of Physics and Astronomy, Stony Brook University, Stony Brook, New York 11794-3800, USA

#These authors contributed equally: Yonghe Liu, Hanqi Pi.

*Correspondence to: quansheng.wu@iphy.ac.cn; yqli@iphy.ac.cn; yang.xu@iphy.ac.cn

## Device fabrication

We fabricate ZrTe$_5$ devices using a modified layer-by-layer dry transfer method, where the van der Waals force between the 2D materials is utilized to pick up and transfer them onto different substrates. The basic fabrication process of a typical device is as follows. Graphite and hexagonal boron nitride (hBN, with typical thickness about 5 - 20 nm) thin flakes are exfoliated from bulk crystals and transferred by a polycarbonate (PC) stamp onto SiO$_2$/Si substrate to form a hBN/graphite stack and serve as the bottom gate. The Ti/Au contacts (~ 8 nm) are pre-patterned into a quasi-Hall bar structure on the hBN/graphite by the e-beam lithography and metallization. The polymer residues on hBN are cleaned using an atomic force microscope (AFM) under the contact mode to achieve high quality of the interface. The rectangular-shaped ZrTe$_5$ (grown by tellurium flux method[1]) flakes are mechanically exfoliated on SiO$_2$/Si substrate in an argon-filled glove box system with both O$_2$ and H$_2$O concentration < 0.1 ppm. The high-quality ZrTe$_5$ flakes are carefully selected by their optical contrast under a microscope to have a thickness of 10 - 30 nm. Another stack of graphite/hBN (as the top gate) on a PC stamp is used to pick up the ZrTe$_5$ flake and then finally released onto the pre-patterned Ti/Au electrodes. The device schematic and optical micrograph is shown in Fig. 1a and 1b, respectively. To minimize the degradation in ambient conditions, wire bonding is also proceeded inside argon-filled glove boxes and the final device, is quickly loaded into the cryostat after removal from the glove box.

## Transport measurements

Electric transport measurements are performed in a helium-4 vapor flow cryostat with variable temperatures of 1.6 - 250 K and magnetic fields up to 9 T. The devices are typically sourced with a constant AC excitation current of 10 - 100 nA. Synchronized NF Corporation LI5640 lock-in amplifiers are used to measure the longitudinal and Hall resistances ($R_{xx}$ and $R_{xy}$) concurrently on ZrTe$_5$ devices. The lock-in amplifiers are operated at low frequencies. Two Keithley 2901 source meters are used to supply the top and bottom gate voltages ($V_{tg}$ and $V_{bg}$) through the graphite gate electrodes.

## Calculations

We perform first-principles calculations using the Vienna ab initio simulation package (VASP)[2] with the generalized gradient approximation of Perdew-Burke-Ernzerhof exchange-correlation potential[3]. Lattice parameters of a = 3.980 Å, b = 14.470 Å, and c = 13.676 Å are used. The self-consistent calculation is carried out on a 11×11×11 k-mesh with the energy cutoff of 500 eV. The temperature-dependent conductivity with fixing carrier concentration is calculated using the Boltztrap code[4]. It processes the single-particle eigenvalues obtained from VASP and carries out on a 41×41×41 k-mesh. To differ the relaxation time of electrons and holes, we modify the source code of BoltzTrap to output the conductivity from every relevant band.

**Magnetoresistance**

We have performed magnetoresistance (MR) measurements at different doping levels and temperatures. The temperature induced evolution of MR (defined as $\frac{R_{sq}(B)-R_{sq}(B=0)}{R_{sq}(B=0)} \times 100\%$) at three representative gate voltages -0.75 V (p-doped), 0.40 V (CNP), and 1.20 V (n-doped) are presented in Fig. S3a. The corresponding Kohler's plot with $\Delta R_{sq}(B)/R_{sq}(0)$ versus $B/R_{sq}(0)$ are shown in Fig. S3b. If there is only one type of scattering at the chemical potential, the temperature-dependent Kohler plots of MR curves should conform to the same functional form[5-7]:

$$\frac{\Delta R_{sq}(B)}{R_{sq}(B=0)} = F\left[\frac{B}{R_{sq}(B=0)}\right] \quad (1)$$

The MR curves in the Kohler's plots do not converge and the difference is large in the electron-doped side, consistent with the multiband charge transport in the ZrTe$_5$ thin device.

**Quantum oscillations**

Here we present experimental signatures of Landau level formation obtained on the same device as the one shown in the main text. We have been able to observe the quantum oscillations in such high-mobility devices. The color contour maps of the 2D longitudinal resistance and Hall resistances as functions of $V_{tg}$ and $B$ field are shown in Fig. S7a-b. Clear Landau levels can be resolved by calculating the longitudinal conductivity $\sigma_{xx} = \rho_{xx}/(\rho_{xx}^2 + \rho_{xy}^2)$ and plotting its derivatives ($\frac{\partial \sigma_{xx}}{\partial V_{tg}}$ and $\frac{\partial \sigma_{xx}}{\partial B}$) in Fig. S7c-d. The quantum oscillations are developed at magnetic fields as low as 1~2 T, in accordance with a quantum mobility of 5,000-10,000 cm$^2$/Vs and close to the field-effect mobilities. These oscillations arise from the band 1 located at the Γ point. We have also traced some of the Landau levels by the dashed curves in Fig. S7c. The Landau fans converge to two different gate voltages at zero $B$ fields for the hole-type and electron-type of charge carriers, which should correspond to the band edges of the conduction and valence bands at the Γ point, respectively. It is also consistent with our assignment of the existence of a small gap discussed in the main text. As multiple bands contribute to the transport in ZrTe$_5$, the oscillation amplitudes are not obvious enough in the MR data.

**Scatterings mechanisms**

Scattering is an important but also complicated problem in general. In realistic devices, multiple scattering mechanisms could contribute to the transport process. Taking the well-known monolayer graphene as an example, there are numerous experimental and theoretical papers studying the underlying scattering mechanisms, the suppression of which have greatly helped improve the quality of graphene devices. The well-known sources of scatterings are electron-phonon scatterings (which contribute to a linear in $T$ resistance above the Bloch-Grüneisen temperature), electron-electron scatterings (Fermi-liquid $T^2$ behavior), Coulomb (long-range) scatterings, lattice disorder (short-range) scatterings, etc. The separation of each contribution is not a trivial task.

So far, there are far fewer papers discussing the scattering mechanisms in detail on ZrTe$_5$. We first attempt to extract the contribution from the electron-phonon and electron-electron scatterings at higher doping levels where clear metallic behavior is observed below $T_p$. As shown below in Fig. S8a-b, we give two examples (one at hole doping with $V_{tg}$ = -0.75 V and the other at electron doping with $V_{tg}$ = 0.95 V) of such fittings. The temperature-dependent resistance $R_{sq}$ is fitted by $A_1T+C_1$ (electron-phonon scatterings, green dashed curves) at elevated temperatures and $A_2T^2+C_2$ (electron-electron scatterings, blue dashed curves) at lower temperature ranges. The fitting parameters $A_1$ and $A_2$ are extracted and plotted as functions of the carrier density. As can be seen, the hole-doped side features nearly density independent $A_1$ and $A_2$, whose values are much larger on the electron-doped. The density independent linear temperature dependences of resistivity have been reported for monolayer graphene[8], where acoustic phonon scatterings contribute to the nearly constant $A_1 = (\frac{h}{e^2}) \frac{\pi^2 D_A^2 k_B}{2h^2 \rho_s v_{ph}^2 v_F^2}$. The $v_F$ is the Fermi velocity, $v_{ph}$ is the velocity of sound, $D_A$ is the acoustic deformation potential, and $\rho_s$ is the mass density. The scattering time $\tau$, which is inversely proportional to $A_1$, differs by a factor 5-10 in the hole- and electron- doped sides, validating the assumption of our theoretical model in simulating the $R$-$T$ curves in Fig. 5c and Fig. S11. The difference could stem from the distinct phonon assisted intervalley scatterings in the conduction and valence bands. The lower-temperature $R$-$T$ follows the Fermi-liquid like $T^2$ behavior, indicating the ZrTe$_5$ a weak interaction system.

While at the lowest temperature when the electron-phonon and electron-electron scatterings are greatly suppressed, the dominant scatterings contribute to the residual resistances are likely from the Coulomb scatterings on charged impurities in the bulk of the material. The full width at half maximum (FWHM) of the resistivity peak at the charge neutrality point (upper panel of Fig. 1d) is related to the amount of charge inhomogeneity $\sim 1 \times 10^{12}$ cm$^{-2}$, consistent with that extracted from the maximum Hall coefficient. Near the charge neutrality, the charge carriers in ZrTe$_5$ break up into electron and hole puddles due to the random

Coulomb potential fluctuations and ineffective Coulomb screenings, resulting in the finite sheet resistance ~4.5 kΩ even when a small gap presents. The sheet conductivity σ$_{xx}$ maintains linearity in the whole gating range, indicating negligible contribution from the short-range scatterings (should come into play at very high densities when Coulomb impurities are highly screened).

To summarize, the current ZrTe$_5$ device quality just matches the early-day graphene on SiO$_2$. There is still a large potential for improvement. Future work is needed for further clarifying the role of intervalley scatterings (that induced by the absorption and emission of large-momentum phonons) and keep working on reducing the Coulomb scatterings for realistic electronic applications.

**Two-band model for fitting the Hall resistances**

We employ the semiclassical two-carrier model, which gives the following equations for the Hall resistance:

$$R_{xy} = \frac{B[B^2\mu_1^2\mu_2^2(n_1+n_2) + n_1\mu_1^2 + n_2\mu_2^2]}{e[B^2\mu_1^2\mu_2^2(n_1+n_2)^2 + (|n_1|\mu_1 + |n_2|\mu_2)^2]} \quad (2)$$

Where $n_{1,2}$ is and $\mu_{1,2}$ are the density and mobility for two types of charge carriers respectively. The total resistance $R_{sq}$ satisfies $R_{sq}(B=0\,\text{T}) = \frac{1}{e(|n_1|\mu_1+|n_2|\mu_2)}$ and $\frac{d\rho_{xy}(B)}{d(B)}|_{B=0\,\text{T}} = \frac{(n_1\mu_1^2+n_2\mu_2^2)}{e(|n_1|\mu_1+|n_2|\mu_2)^2}$, posing an additional constraint on fitting parameters, where $e$ is the elementary charge. From the equation (2), the magnetic-field-dependent Hall coefficient has three limits[9]:

$$R_H = \begin{cases} \frac{(n_1\mu_1^2+n_2\mu_2^2)}{e(|n_1\mu_1|+|n_2\mu_2|)^2}, & (\mu_1 B \ll 1 \text{ and } \mu_2 B \ll 1) \quad (3) \\ \frac{1}{e(n_1+n_2)}, & (\mu_1 B \gg 1 \text{ and } \mu_2 B \gg 1) \quad (4) \\ \frac{1}{e}\left(\left(n_1 + \mu_1\mu_2\frac{n_2^2 B^2}{n_1}\right)^{-1}\right), & (\mu_1 B \gg 1 \text{ and } \mu_2 B \ll 1) \quad (5) \end{cases}$$

From the equation (3) and (4), the Hall coefficient at low magnet field is defined as $R_H^l$, which is dominated by the high mobility carriers. While the Hall coefficient at high magnet field is defined $R_H^h$, which reflects the total carrier density.

## Temperature-dependent resistivity from a simplified model

To confirm our theory about the peak of the temperature-dependent resistance of ZrTe$_5$, we construct a model considering a simple band structure with only one conduction band and valence band separated by a narrow gap (Fig. S9). The dispersion relations as represented below,

$$\begin{cases} E_c = \dfrac{\hbar^2 k^2}{2m_e} + \dfrac{\Delta}{2}, \\ E_v = -\dfrac{\hbar^2 k^2}{2m_h} - \dfrac{\Delta}{2}, \end{cases}$$

where $\Delta = 0.1$ eV is the energy gap, $m_e = 1$ ($m_h = 0.8$) is the effective mass of electrons (holes). For simplicity, we let $\hbar = V/\pi^2 = e = 1$ in the following calculation.

The density of states (DOS) of the conduction band $g_e(\varepsilon)$ and valence band $g_h(\varepsilon)$ are

$$g_e(\varepsilon) = \frac{V}{2\pi^2}\left(\frac{2m_e}{\hbar^2}\right)^{\frac{3}{2}}\sqrt{\varepsilon - \frac{\Delta}{2}},$$

$$g_h(\varepsilon) = \frac{V}{2\pi^2}\left(\frac{2m_h}{\hbar^2}\right)^{\frac{3}{2}}\sqrt{-\varepsilon - \frac{\Delta}{2}}.$$

DOS above the Fermi level is obviously larger than that below since $m_e > m_h$. When the chemical potential at $T = 0$ K is set to $\mu$, the charge carrier concentration is

$$n_0 = \frac{V}{3\pi^2}\left(\frac{2m_e}{\hbar^2}\right)^{\frac{3}{2}}\left(\mu - \frac{\Delta}{2}\right)^{\frac{3}{2}}, \qquad \text{when } \mu > \Delta/2,$$

$$n_0 = -\frac{V}{3\pi^2}\left(\frac{2m_h}{\hbar^2}\right)^{\frac{3}{2}}\left(-\mu - \frac{\Delta}{2}\right)^{\frac{3}{2}}, \qquad \text{when } \mu < \Delta/2.$$

We obtain the chemical potential at different temperature $\tilde{\mu}(T)$ by fixing the charge carrier concentration $n_0 = \tilde{n}_e(T) - \tilde{n}_h(T)$. The electron concentration $\tilde{n}_e(T)$ and hole concentration $\tilde{n}_h(T)$ at different temperatures are:

$$\tilde{n}_e(T) = \int_{\frac{\Delta}{2}}^{+\infty} g_e(\varepsilon)f(\varepsilon - \tilde{\mu})d\varepsilon,$$

$$\tilde{n}_h(T) = \int_{-\infty}^{-\frac{\Delta}{2}} g_h(\varepsilon)[1 - f(\varepsilon - \tilde{\mu})]d\varepsilon,$$

$$= \int_{\frac{\Delta}{2}}^{+\infty} g_h(\varepsilon)f(\varepsilon + \tilde{\mu})d\varepsilon,$$

where $f$ is the Fermi-Dirac distribution function. The relevant quantities are shown in Fig. S10, including the carrier concentration $n_0$ at different chemical potential $\mu$, temperature-dependent chemical potential $\tilde{\mu}(T)$, electron concentration $\tilde{n}_e(T)$ and hole concentration $\tilde{n}_h(T)$. Due to the higher DOS of the conduction band, not only the electron pocket changes more significantly than the hole pocket when tuning the chemical potential, but the change of $\tilde{\mu}(T)$ with $\mu > 0$ is greater than that with $\mu < 0$, as represented in Fig. S10a-b. Fig. S10c-d show that there is only one kind of charge carrier at low temperatures with constant concentration. However, as the temperature rises, the other type of charge carrier appears at $T_p$ and both of them grow with the temperature.

Assuming the relaxation time of electrons and holes are the same and inversely proportional to the temperature, $\tau_e = \tau_h = \frac{1}{T}$, we obtain the temperature-dependent resistance $\rho(T) = \frac{1}{e}\frac{1}{\tilde{n}_e(T)\mu_e(T)+\tilde{n}_h(T)\mu_h(T)}$ at different chemical potential $\mu$ as shown in Fig. S11a, where $\mu_e = \frac{e\tau_e}{m_e}$ and $\mu_h = \frac{e\tau_h}{m_h}$ are the mobility of electrons and holes, respectively. The resistance increases linearly at low temperatures because electron and hole concentrations remain constant and the mobilities decrease as proportional to $1/T$. By increasing the temperature, the minority carriers emerge such that $\tilde{n} = \tilde{n}_e(T) + \tilde{n}_h(T)$ increases. Eventually, the slope $d\rho/dT$ decreases and the resistance reaches its maxima at $T_p$ when the growth rate of electron(hole) concentration equals the decreasing rate of mobility.

The peak temperatures ($T_p$) are extracted to construct a phase diagram as illustrated in Fig. S11d. It is found that $T_p$ increases with the magnitude of charge carrier density $|n_0|$, resulting from the higher $T_c$ that the minority carriers appear as shown in Fig. S10c-d. Moreover, $T_p$ in the electron-doped region is lower than the hole-doped region when $|n_0|$ is the same. As $T_p$ is determined by the increasing concentration $n_{e,h}(T)$ and the decreasing mobility $\mu_{e,h}(T)$, we can analyze the asymmetry of the phase diagram from the two aspects. Due to the larger DOS of the conduction band, the change of $n_{e,h}(T)$ in the electron-doped region compared to the hole-doped region with the same $|n_0|$ is more drastic. Besides, the magnitude and the rate of change of $\mu_h(T)$ are larger than $\mu_e(T)$ because of the smaller effective mass. As the resistance is mainly determined by the majority carriers, the increasing rate of $n_h(T)$ is supposed to surpass the decreasing rate of $\mu_h(T)$ at a higher $T_p$ in the hole-doped region, while $T_p$ determined by $n_e(T)$ and $\mu_e(T)$ in the electron-doped region is lower.

When increasing the relaxation time of the hole carriers by a factor of 5, i.e., $\tau_h = 5\tau_e = \frac{5}{T}$, the temperature-dependent resistance $\rho(T)$ and the phase diagram are illustrated in Fig. S11b and S11e. Obviously, the intensity of $\rho(T)$ is reduced due to the increasing $\mu_h(T)$. Because the type of majority carriers is different, the reduction of $\rho(T)$ in the hole-doped region is larger than that in the electron-doped region as shown in Fig. S11b. For the same reason, the effect of increasing $\mu_h(T)$ on $T_p$ in the hole-doped region is opposite to that in the electron-doped region, which further exacerbates the asymmetry of the $T_p$ dispersion as shown in Fig. S11e. If we continue to increase the relaxation time of the hole carriers by a factor of 10, the above phenomena are more prominent as shown in Fig. S11c and S11f.

**Supplementary Figure**

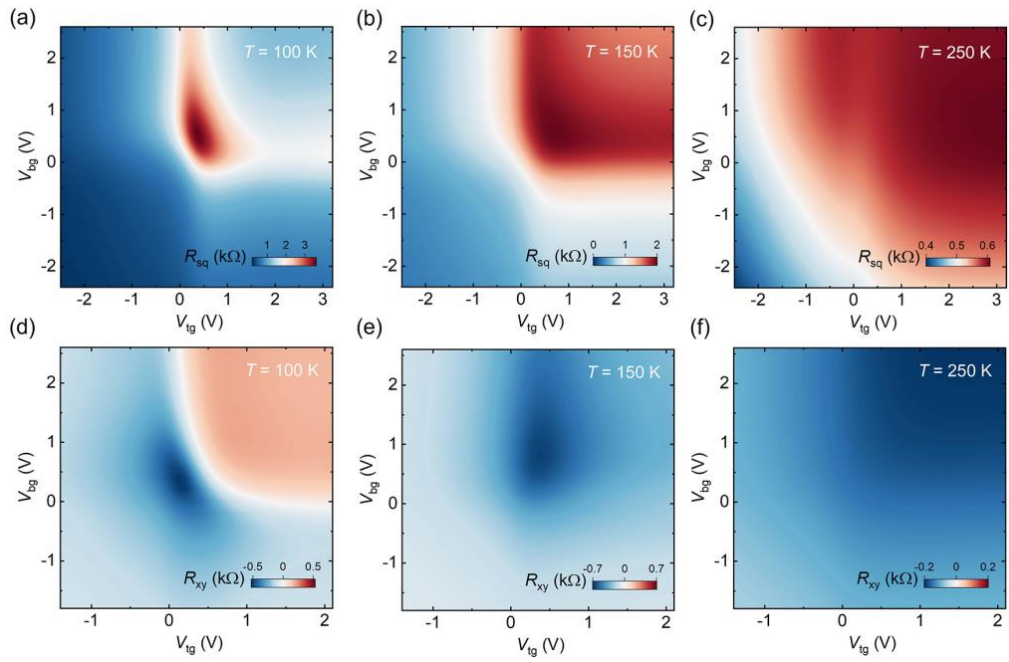

**Fig. S1. 2D color maps of $R_{sq}(B = 0$ T) and $R_{xy}(B = 1$ T) as a function of dual-gated voltages at three higher temperatures**. (a-c) The longitudinal sheet resistance $R_{sq}$ ($B = 0$ T) and (d-f) Hall resistance $R_{xy}$ (anti-symmetrized at $B = \pm 1$ T) as functions of $V_{tg}$ and $V_{bg}$ measured at $T = 100, 150, 250$ K, respectively. The charge transport in the entire gating range is mainly hole-like above ~150 K.

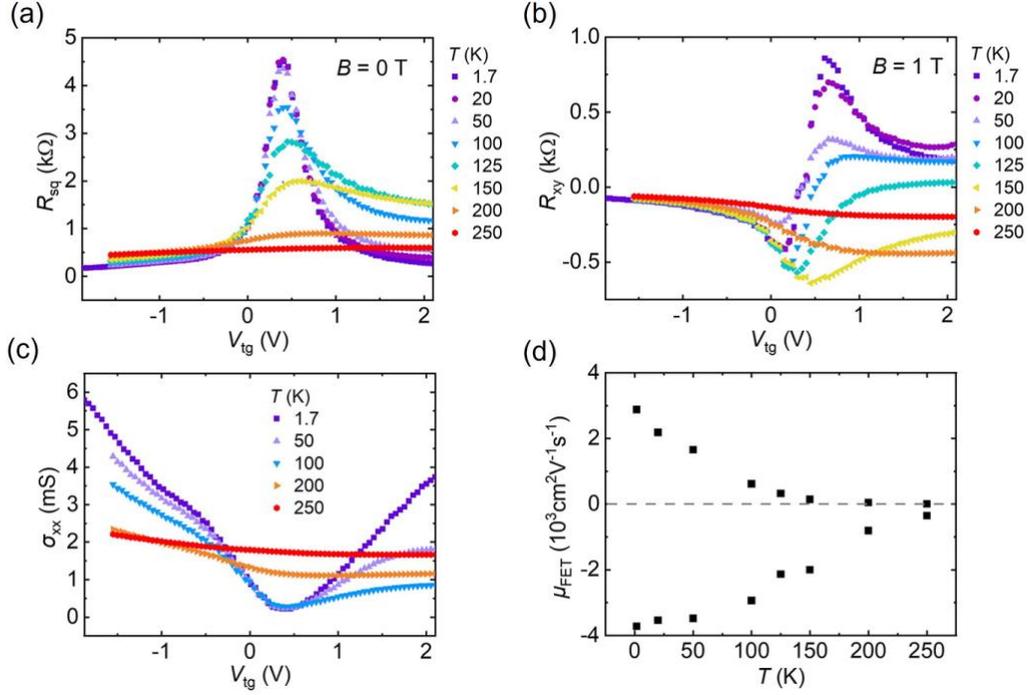

**Fig. S2. Temperature dependence of the field effect at zero and low magnetic field.** (a-c) The longitudinal sheet resistance $R_{\rm sq}$ ($B = 0$ T), Hall resistance $R_{\rm xy}$ (antisymmetrized at $B = \pm 1$ T), and the longitudinal conductance $\sigma_{\rm xx}$ as functions of the gate voltage at a few temperatures. Below 150 K, the Hall resistance $R_{\rm xy}$ shows a sign change via tuning the gate voltage, whereas it becomes all negative (holes dominate charge transport) above 150 K. (d) The field effect mobility extracted by $\mu_{\rm FET} = \frac{d\sigma_{\rm xx}}{C_{\rm tg}dV_{\rm tg}+C_{\rm bg}dV_{\rm bg}}$, where $C_{\rm tg}$ and $C_{\rm bg}$ are the parallel-plate capacitances of the top and bottom gates. The $\mu_{\rm FET}$ is negative for holes and positive for electrons. The holes tend to have higher mobilities at higher temperatures.

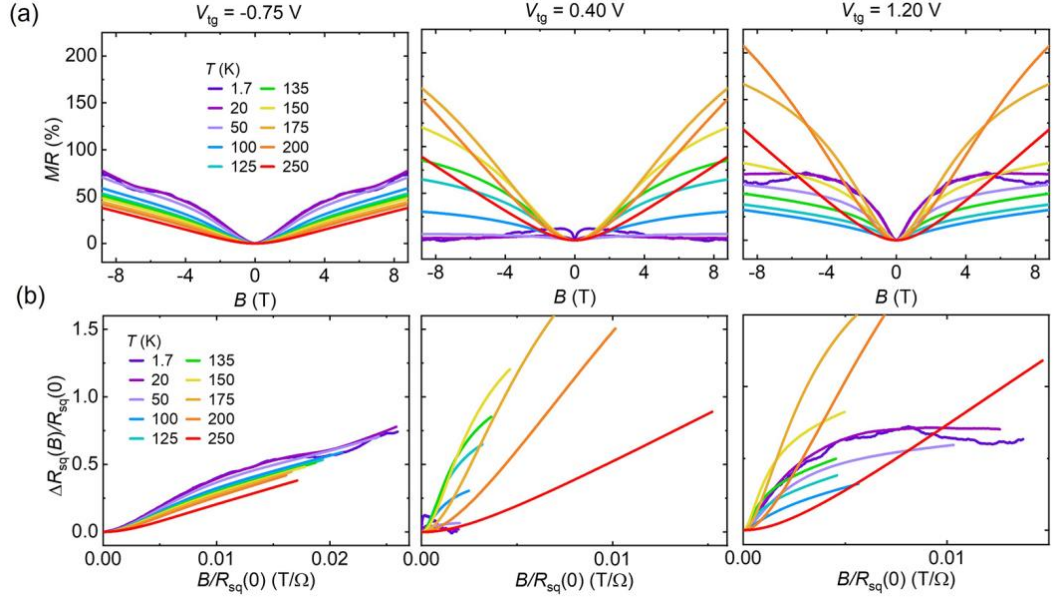

**Fig. S3. Temperature dependence of the magnetoresistance studied at three representative gate voltages.** (a, b) Temperature-dependent evolution (from 1.7 to 250 K) of the magnetoresistance (MR) and the corresponding Kohler's plots shown at $V_{tg}$ = -0.75 V (hole-doped side), 0.40 V (charge neutrality), and 1.20 V (electron-doped side), respectively.

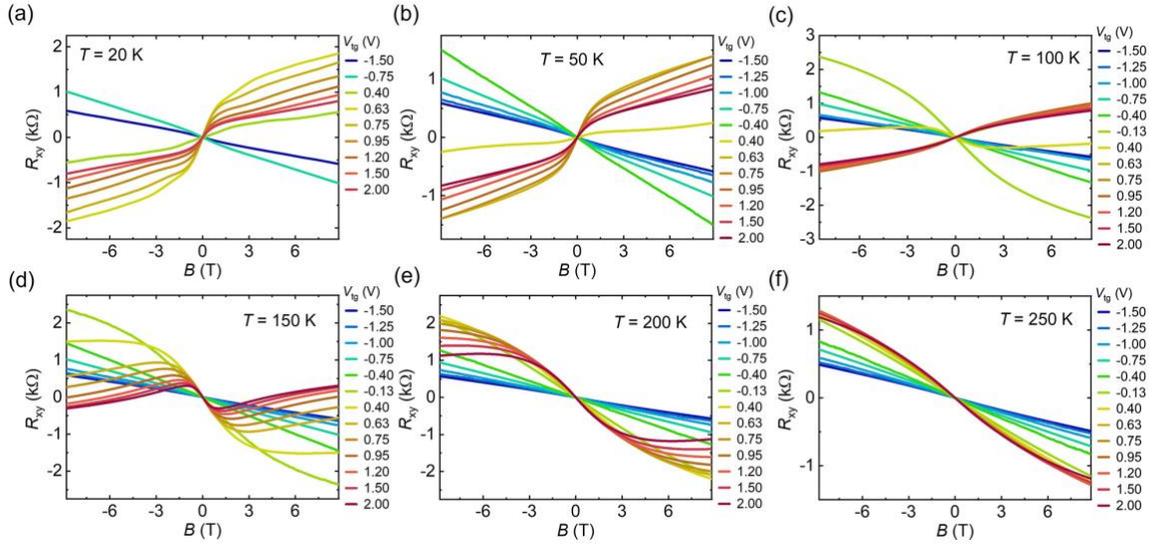

**Fig. S4. The magnetic field dependences of the Hall resistance at different gate voltages and temperatures.**

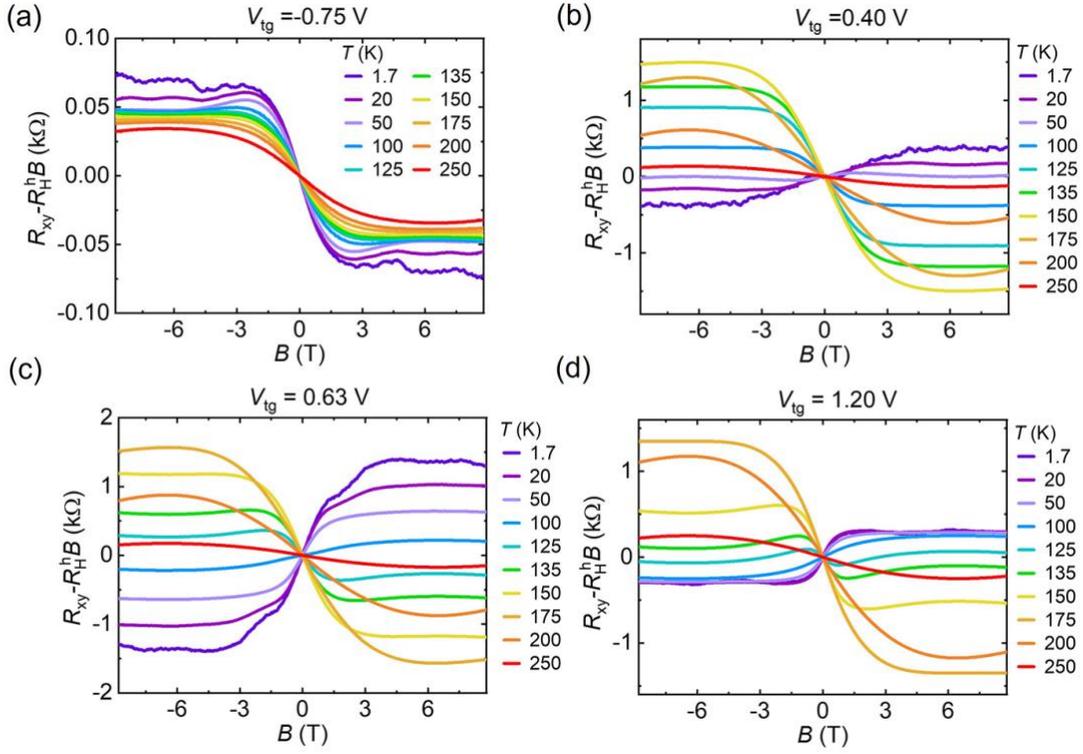

**Fig. S5. Multiband induced anomalous-Hall-like response at different temperatures.** (a-d) The extracted $(R_{xy} - R_H^h B)$ as a function of magnetic field at different temperature for four representative gate voltages: $V_{tg}$ = -0.75, 0.40, 0.63, 1.20 V, respectively. The "false" saturation field $B_0$ (defined in the main text) is always less than 4 T.

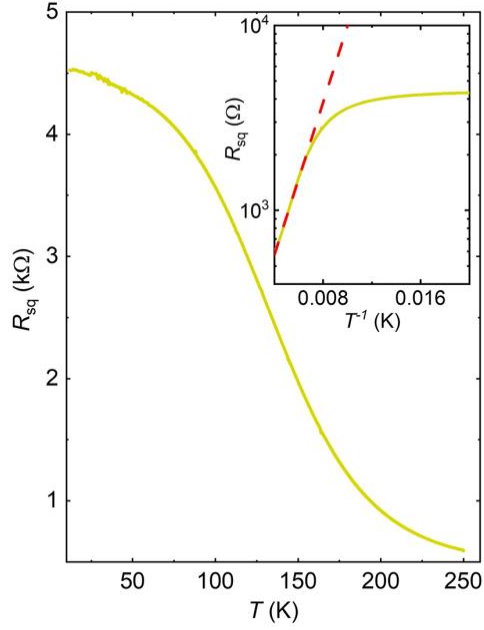

**Fig. S6. The insulating behavior of $R_{sq}(T)$ at CNP.** The resistance $R_{sq}(T)$ exhibits insulating behaviors ($\frac{dR_{sq}}{dT} < 0$) upon cooling to the lowest temperature. Inset shows an activation gap $\Delta \sim 40$ meV extracted from the Arrhenius plot of $R_{sq} \sim \exp(\frac{\Delta}{k_B T})$.

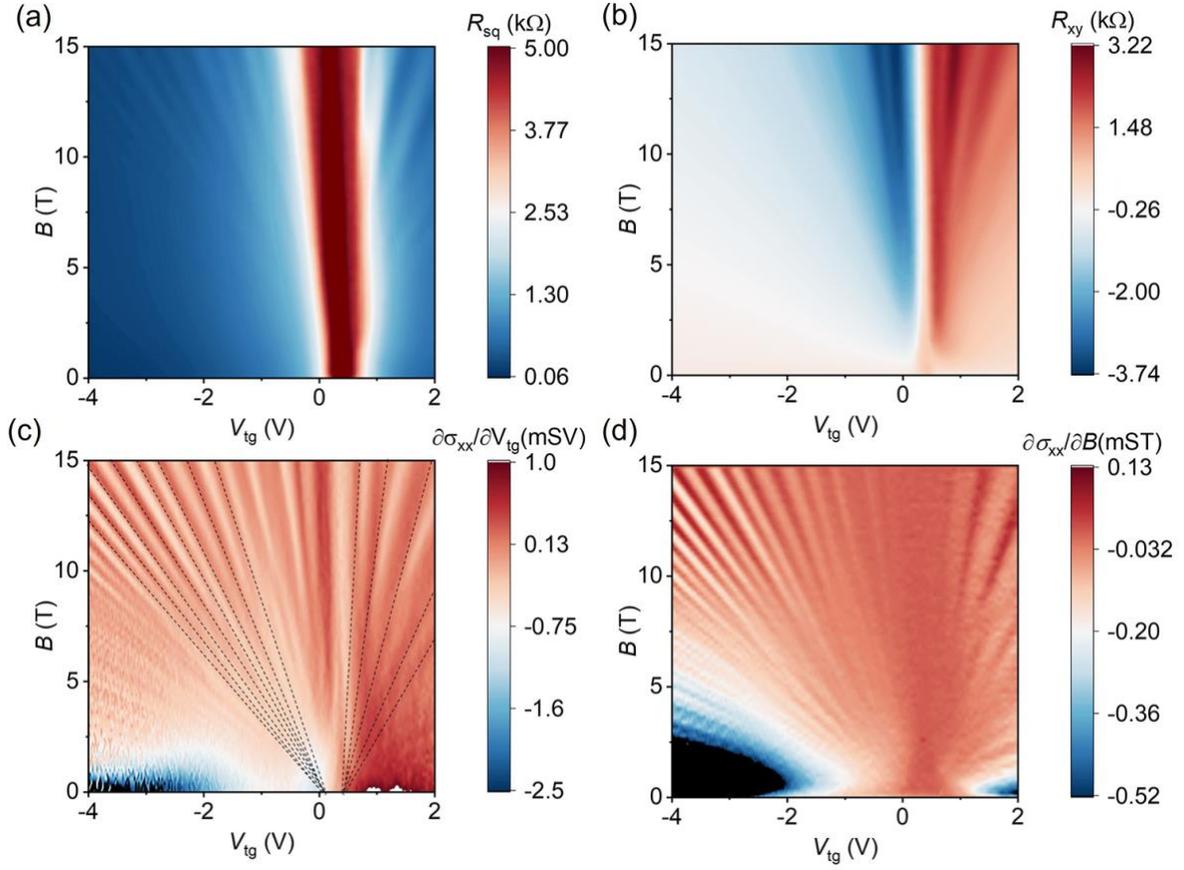

**Fig. S7. Landau fan measurements.** (a-d) Color contour plots of the longitudinal sheet resistance $R_{sq}$ (a), Hall resistance $R_{xy}$ (b), and the $\sigma_{xx}$ derivatives [$\frac{\partial \sigma_{xx}}{\partial V_{tg}}$ (c) and $\frac{\partial \sigma_{xx}}{\partial B}$ (d)] as functions of $V_{tg}$ and magnetic field $B$. Landau levels can be resolved more clearly in (c-d). The measurements are performed at $T = 2$ K and $V_{bg} = 0.4$ V.

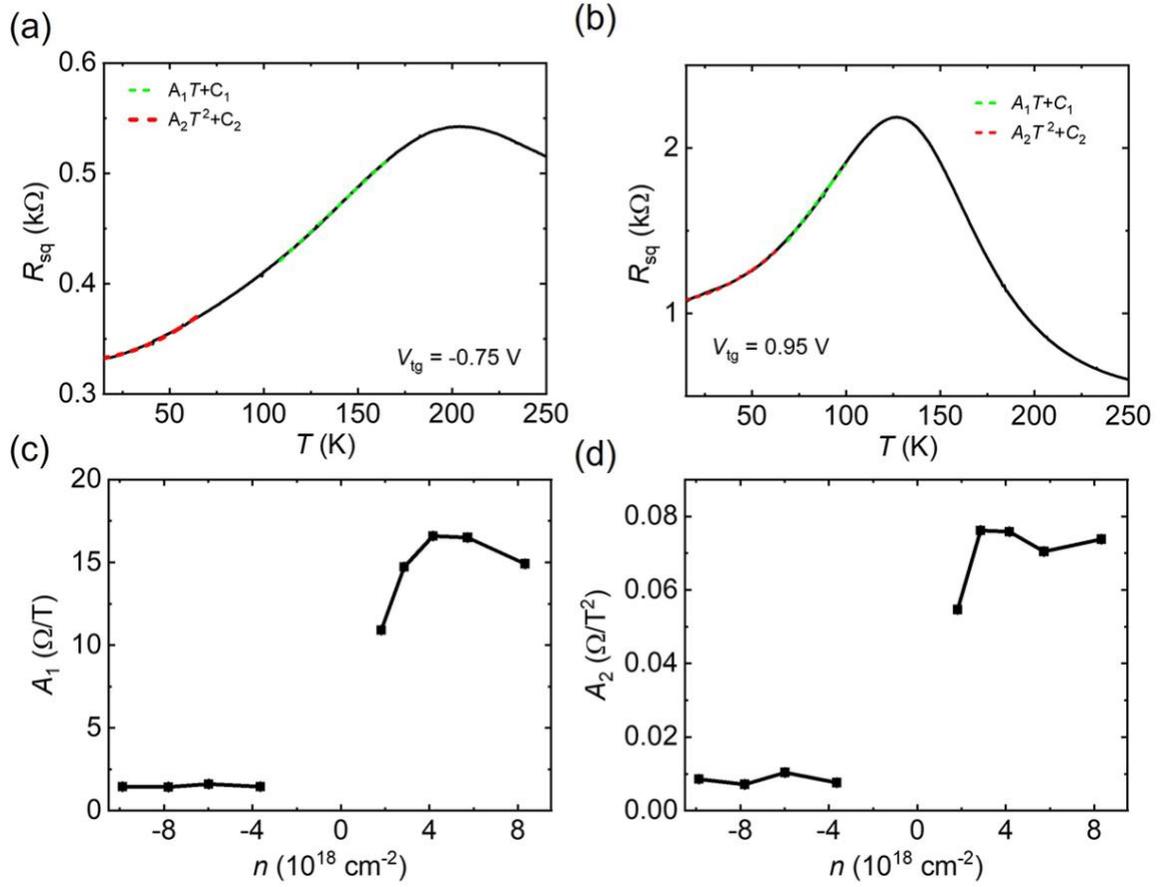

**Fig. S8. Electron-phonon and electron-electron scatterings.** (a-b) The $R_{sq}(T)$ curves at $V_{tg}$ = -0.75 V (a) and 0.95 V (b). The $R_{sq}(T)$ are fitted by $A_1T+C_1$ (electron-phonon scatterings, green dashed curves) and $A_2T^2+C_2$ (electron-electron scatterings, blue dashed curves) at different temperature ranges. (c-d) The fitting parameters $A_1$ and $A_2$ as functions of the carrier density, respectively.

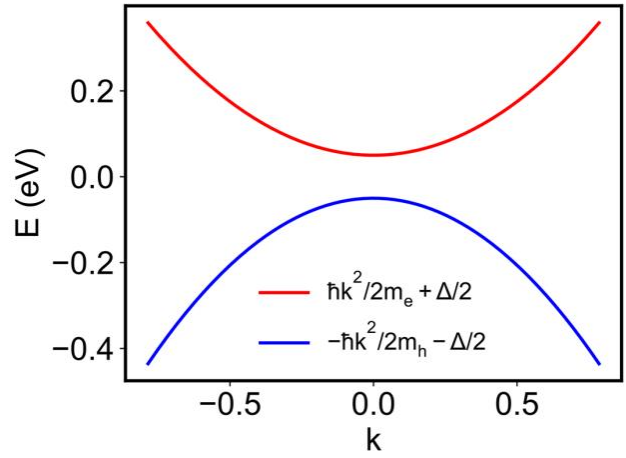

**Fig. S9. The energy dispersion relation of the narrow-gap semiconductor model.**

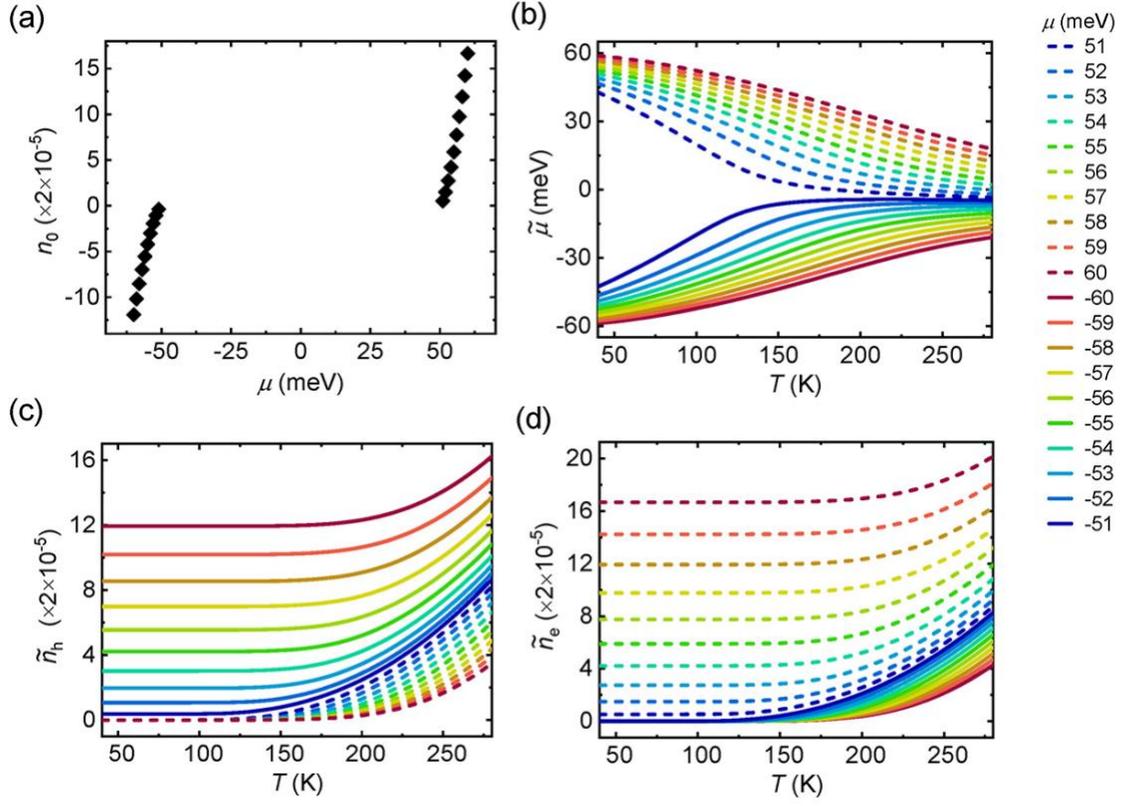

**Fig. S10. Exacted parameters and evolution with temperature.** (a) Carrier concentration $n_0$ at different chemical potential $\mu$ (zero temperature), (b-d) temperature-dependent chemical potential $\tilde{\mu}(T)$ (b), electron concentration $\tilde{n}_e(T)$ (c) and hole concentration $\tilde{n}_h(T)$ (d).

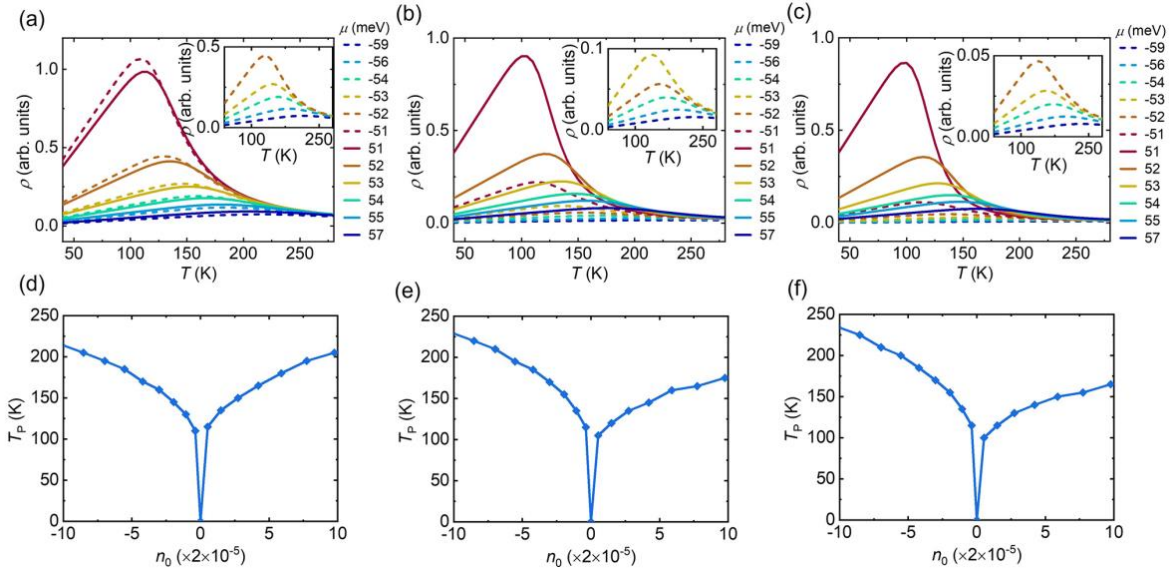

**Fig. S11. The evolution of resistivity $\rho(T)$ and the corresponding $T_\text{P}(n_0)$ with increasing scattering rate ratio from the simplified narrow-gap semiconductor model.** (a-c) The resistivity $\rho$ as a function of temperature for several carrier densities $n$. The inset shows a zoom-in plot of several most p-doped curves. The resistivity $\rho$ is equal to $\frac{1}{\sigma}$, where $\sigma$ is the conductivity and the $\frac{\sigma}{\tau_{h(e)}}$ is obtained directly from the band structure ($\tau_{h(e)}$ is the scattering time for holes (electrons)). The results with $\frac{\tau_h}{\tau_e}$ assumed to be 1, 5, 10 are shown in (a)-(c), respectively. (e-f) The corresponding $T_\text{P}$ plotted as a function of carrier density $n_0$. As the rate $\frac{\tau_h}{\tau_e}$ increases, the asymmetry of $T_\text{P}(n_0)$ becomes more obvious.

# Reference for SI